\documentclass[preprint,aps,showpacs,showkeys,dvips]{revtex4}

\usepackage{graphics}
\usepackage{dcolumn}
\usepackage{bm}

\begin{document}

\title{Anomalous Single Production of the Fourth Generation Neutrino at Future ep Colliders}

\author{ A. K. \c{C}ift\c{c}i}
\email{ciftci@science.ankara.edu.tr}
\affiliation{Physics Department, Faculty of Sciences, Ankara University, 06100 Tandogan,
Ankara, Turkey}

\author{R. \c{C}ift\c{c}i}
\email{rena.ciftci@gmail.com}
\affiliation{Dept. of Eng. of Physics, Faculty of Eng., Ankara University, 06100 Tandogan,
Ankara, Turkey}

\author{S. Sultansoy}
\email{ssultansoy@etu.edu.tr}
\affiliation{Physics Section, Faculty of Sciences and Arts, TOBB University of Economics and Technology,
Ankara, Turkey}
\altaffiliation[Also at ]{Institute of Physics, Academy of Sciences, H. Cavid Avenue 33, 
Baku, Azerbaijan}

\date{\today}

\begin{abstract}
Possible single productions of the fourth standard model generation neutrino via anomalous interactions at the future ep colliders are studied. Signatures of such anomalous processes and backgrounds are discussed in detail. Discovery limits for neutrino mass and achievable values of anomalous coupling strength are determined.
\end{abstract}
\maketitle
\section{Introduction}	

It is known that the Standard Model (SM) does not predict number of fundamental fermion generations. This number is restricted from below with LEP I data on invisible decays of Z boson as $n_g\geq3$ {[}1{]}. On the other hand, $n_g<9$ from asymptotic freedom of QCD. According to recent precision electroweak data, existence of three or four SM generations is at the same status {[}2-4{]}.

The flavor democracy is a natural hypothesis in the framework of SM as well as a number of models dealing with new physics (see review {[}5{]} and references therein). Concerning Standard Model, flavor democracy predicts the existence of a heavy fourth SM generation {[}6-8{]}. The Dirac masses of the new fermions are predicted to be almost degenerate and lie between 300 and 700 GeV, whereas, the masses of known fermions belonging to lighter three generations appear due to small deviations of the democracy {[}9-11{]}. The quark masses and CKM matrix are given in {[}9, 10{]}. Ref. {[}11{]} gives both masses and CKM matrix (MNS matrix for leptons) for both quarks and leptons.

Obviously, TeV energy colliders are needed for discovery of the fourth SM generation fermions. The fourth generation quarks will be produced in pairs copiously at the Large Hadron Collider (LHC) {[}12, 13{]}. Recently, this process is proposed as the best scenario (after Higgs) for discovery at the LHC {[}14-16{]}. Linear lepton colliders are the best place for pair production of the fourth generation charged lepton and neutrino {[}11, 17, 18{]}. However, discovery limits for pair production at lepton colliders are $2m<\sqrt{s}$. For example, International Linear Collider (ILC) with 500 GeV center of mass energy will cover $m<250$ GeV. The discovery capacity of lepton collider could be enlarged if the anomalous interactions of the fourth generation fermions with the first three ones exist. Such anomalous interactions seems to be quite natural due to large masses of the fourth generation fermions (see argumentation for anomalous interaction for t quark presented in ref. {[}19{]}). These anomalous interactions could provide also single production of the fourth generation fermions at future lepton-hadron colliders (see review {[}20{] and references therein). Depending on the center of mass energy lepton-hadron colliders are named QCD Explorer with $\sqrt{s}$ = 1.4 TeV and Energy Frontier ep collider with $\sqrt{s}$ = 3.74 TeV {[}20-27{]}.

Recently, anomalous production of the fourth generation charged lepton and neutrino at future ep colliders is considered in {[}28{]} and {[}29{]}, respectively. Unfortunately results of the latter one are erroneous due to the wrong Lagrangian for  $\nu_{4} e W$ interactions. Therefore, anomalous production of the fourth generation neutrino at future ep colliders should be reconsidered. This is the aim of this study.

\section{Anomalous Interactions of the Fourth SM Generation Neutrino}

The charged current Lagrangian for SM and the anomalous interactions of the fourth generation neutrino can be rewritten from {[}30, 31{]} with minor modifications as:

\begin{equation}
L_{cc}=\left(\frac{g_{W} }{\sqrt{2}}\right)\overline{l_{i}}\left[\left|V_{\nu_4l_{i}}\right|\gamma_\mu+\frac{i}{2 \Lambda}\kappa^{\nu_{4}l_{i}}_{W}\sigma_{\mu\nu}q^{\nu}\right]P_L\nu_{4}W^{\mu}+h.c. \hspace{2mm}  ,  \hspace{3mm} (\textit{i}=1,2,3)
\end{equation}
The main error of corresponding Lagrangian in {[}29{]} is absence of the MNS matrix element $|V_{\nu_4l_i}|$. Furthermore, the neutral current Lagrangian for the anomalous interactions of the fourth generation neutrino is

\begin{equation}
L_{nc}=\left(\frac{g_{Z}}{2}\right)\overline{\nu_{i}}\frac{i}{2 \Lambda}\kappa^{\nu_{4}\nu_{i}}_{Z}\sigma_{\mu\nu}q^{\nu}P_L\nu_{4}Z^{\mu}+h.c. \hspace{2mm}  ,  \hspace{3mm} (\textit{i}=1,2,3) 
\end{equation}
In Eq's (1) and (2), $\kappa^{\nu_{4}l_{i}}_{W}$ and $\kappa^{\nu_{4}\nu_{i}}_{Z}$ are the anomalous couplings for the charged and neutral currents with a W boson and a Z boson, respectively (in numerical calculations, we suppose $\kappa^{\nu_{4}l_{i}}_{W}=\kappa^{\nu_{4}\nu_{i}}_{Z}=\kappa$). $\Lambda$ is the cutoff scale for the new physics and  $P_L $ is the left handed projection operator; $g_{W}$ and $g_{Z}$ are the electroweak coupling constants. In the above equations
$\sigma_{\mu\nu} = i(\gamma_{\mu}\gamma_{\nu}-\gamma_{\nu}\gamma_{\mu})/2$. 
\begin{figure}
\includegraphics{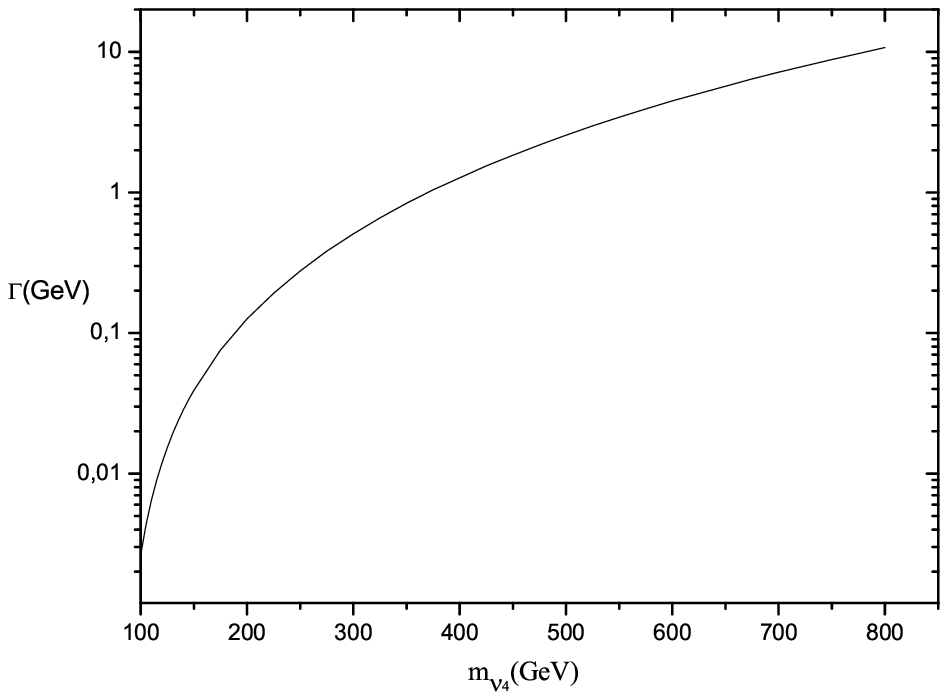}
\includegraphics{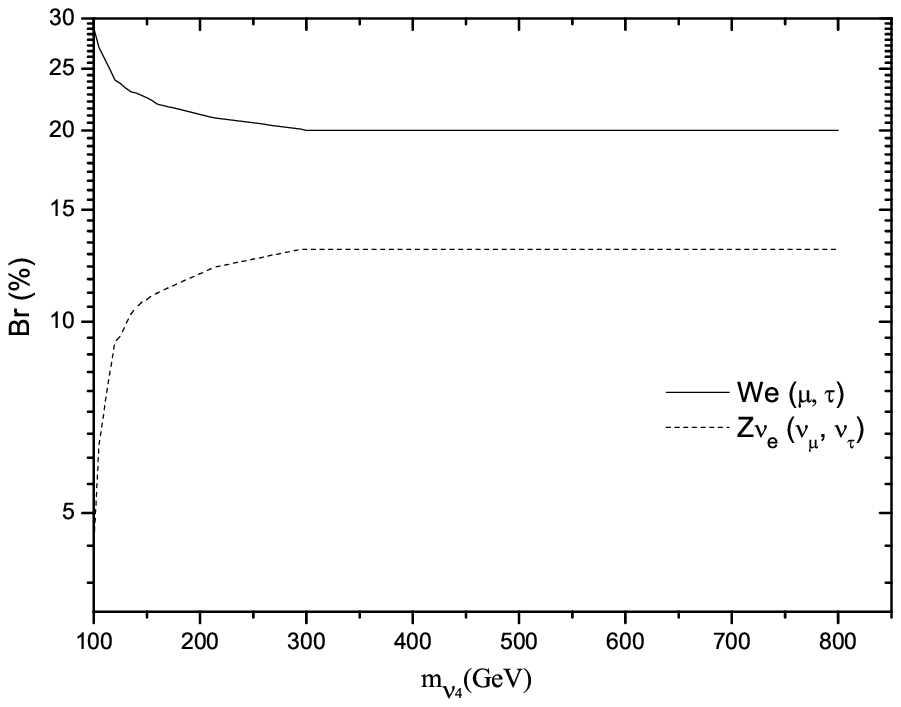}
\caption{(a) The total decay width $\Gamma$ in GeV of the fourth family neutrino and (b) the branching ratios (\%) depending on the mass of the fourth generation neutrino. \protect\label{fig1}}
\end{figure}

\begin{table}
\caption{Branching ratios and total decay widths for $m_{\nu_4}$(GeV).}
\begin{ruledtabular}
\begin{tabular}{ccccc}
$m_{{\nu}_4}$ & $W e^{-}({\mu}^{-},{\tau}^{-})$ & $Z \nu_e(\nu_\mu,\nu_\tau)$ &  $\Gamma_{Tot}$(GeV) \\ 
\hline
100\hphantom{00} & \hphantom{0}29 & \hphantom{0}4 & 0.0026 \\
150\hphantom{00} & \hphantom{0}22 & \hphantom{0}11  & 0.039 \\
200\hphantom{00} & \hphantom{0}21 & \hphantom{0}12  &  0.126 \\
300\hphantom{00} & \hphantom{0}20 & \hphantom{0}13  & 0.508 \\
700\hphantom{00} & \hphantom{0}20 & \hphantom{0}13  & 7.160 \\ 
\end{tabular}
\end{ruledtabular}
\end{table}

Obviously new interactions will lead to additional decay channels of the fourth family neutrino in addition to enhancement of some SM decay channels. In order to compute decay widths, we have implemented the new interaction vertices into the CompHEP {[}32{]}. Results of the calculations for different decay channels of $\nu_4$ assuming  $(\kappa/\Lambda)$ = 1 TeV$^{-1}$ are given in Table I. Experimental upper limit for $|V_{\nu_4l_i}|$ is 0.02 [4]. Therefore, while calculating values in Table I, we have used $|V_{\nu_4l_i}|=0.02$. The total decay width $\Gamma$ of the fourth generation neutrino and the relative branching ratios are plotted in Fig. 1. More realistic values for MNS matrix elements can be taken from ref. {[}17{]}, namely: $V_{\nu_4\tau}$ = 2.34$\times10^{-5}$, $V_{{\nu_4}\mu}$ = 6.81$\times10^{-4}$, $V_{{\nu_4}e}$ = 4.64$\times10^{-4}$. Consequently, we have used these values at rest of our calculations (note that the values in Table I as well as Figure 1 practically do not change).

\begin{table}
\caption{Event numbers of $ep\rightarrow\nu_4X\rightarrow \mu W X$ 
for $\sqrt{s}$ = 1.4 TeV, $(\kappa/\Lambda)$ = 1TeV$^{-1}$}
\begin{ruledtabular}
\begin{tabular}{ccc}
 &\multicolumn{2}{c} {$N_S$}  \\
\cline{2-3}
$m_{\nu_4}$(GeV) & $L_{int}=1$fb$^{-1}$ & $L_{int}=10$fb$^{-1}$ \\ 
\hline
200\hphantom{00} & \hphantom{0}201 & 2010 \\
300\hphantom{00} & \hphantom{0}148 & 1480 \\
400\hphantom{00} & \hphantom{0}106 & 1060 \\
500\hphantom{00} & \hphantom{0}74 & 740 \\
600\hphantom{00} & \hphantom{0}47 & 470 \\
700\hphantom{00} & \hphantom{0}27 & 270 \\
800\hphantom{00} & \hphantom{0}14 & 140  \\
900\hphantom{00} & \hphantom{0}6 & 60  \\
1000\hphantom{00} & \hphantom{0}2 & 22 \\ 

\end{tabular}
\end{ruledtabular}
\end{table}

\begin{table}
\caption{Event numbers of $ep\rightarrow\nu_4X\rightarrow \mu W X$ 
for $\sqrt{s}$ = 3.74 TeV, $(\kappa/\Lambda)$ = 1TeV$^{-1}$}
\begin{ruledtabular}
\begin{tabular}{ccc} 
 &\multicolumn{2}{c} {$N_S$}  \\
\cline{2-3}
$m_{\nu_4}$(GeV)  & $L_{int}=100$pb$^{-1}$ & $L_{int}=1$fb$^{-1}$ \\ 
\hline
200\hphantom{00}  & \hphantom{0}74.3 & 743 \\
300\hphantom{00} & \hphantom{0}60.3 & 603 \\
400\hphantom{00}  & \hphantom{0}50.4 & 504 \\
500\hphantom{00} & \hphantom{0}42.7 & 427 \\
600\hphantom{00} & \hphantom{0}36.4 & 364 \\
700\hphantom{00} & \hphantom{0}30.7 & 307 \\
800\hphantom{00} & \hphantom{0}25.9 & 259 \\
900\hphantom{00} & \hphantom{0}21.4 & 214 \\
1000\hphantom{00} & \hphantom{0}20.1 & 201 \\
1200\hphantom{00} & \hphantom{0}13.8 & 138 \\
1400\hphantom{00} & \hphantom{0}8.8 & 88  \\
1600\hphantom{00} & \hphantom{0}5.3 & 53  \\
1800\hphantom{00} & \hphantom{0}2.9 & 29 \\ 
2000\hphantom{00} & \hphantom{0}1.5 & 15 \\
2200\hphantom{00} & \hphantom{0}0.7 & 7 \\
2400\hphantom{00} & \hphantom{0}0.3 & 3 \\
2600\hphantom{00} & \hphantom{0}0.1 & 1  \\

\end{tabular}
\end{ruledtabular}
\end{table}

\begin{figure}
\includegraphics{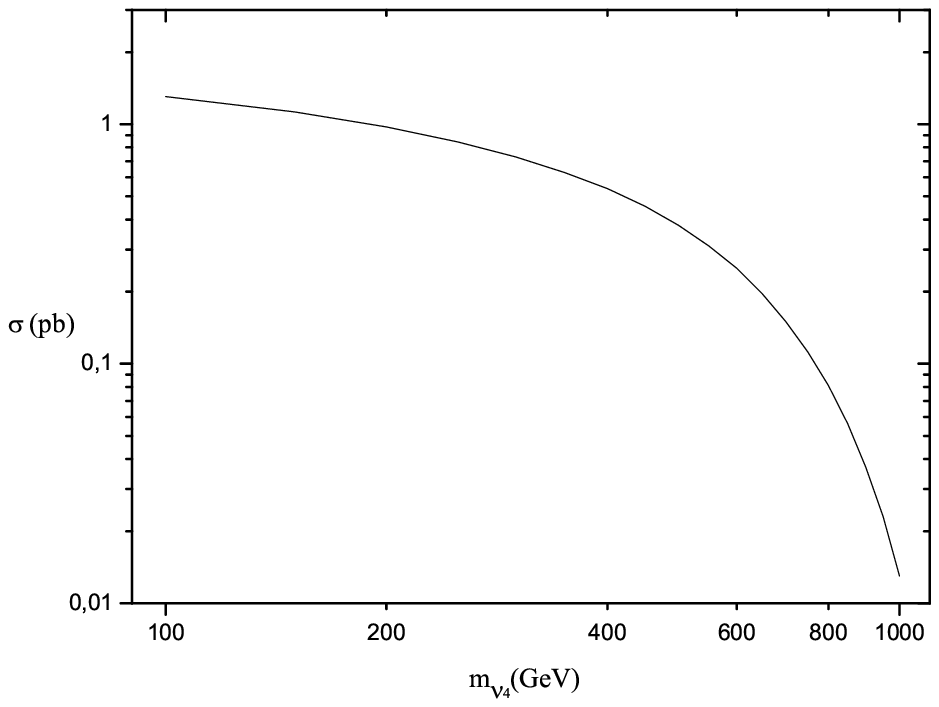}

\caption{\label{fig2}The total production cross-section of the process $ep\rightarrow \nu_4$X as a function of $m_{\nu_4}$ with the center of mass energy $\sqrt{s}$ = 1.4 TeV.}
\end{figure}

\begin{figure}
\includegraphics{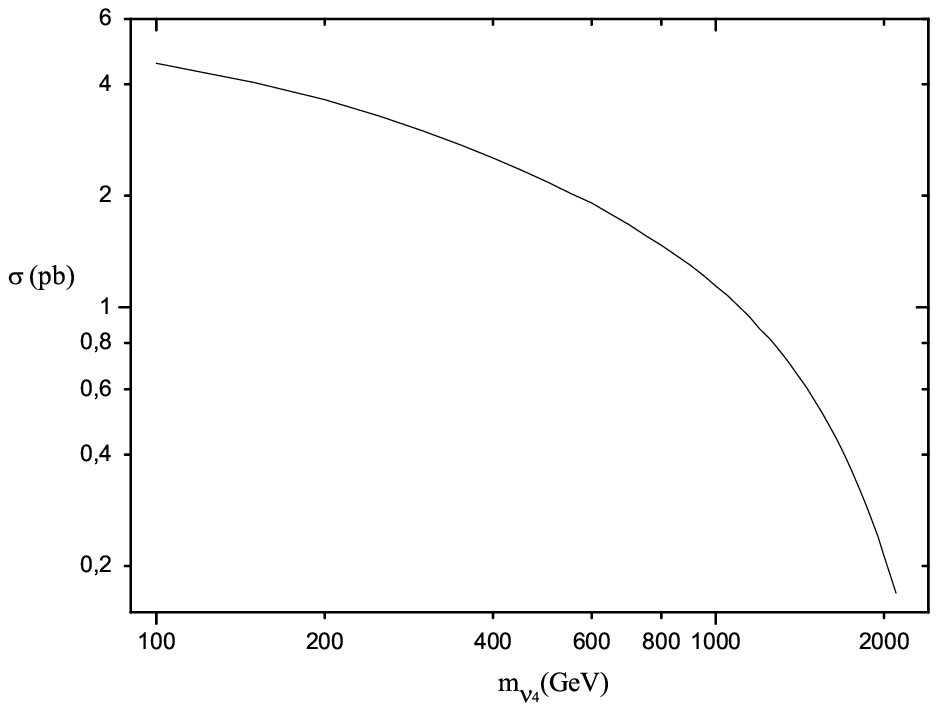}
\caption{\label{fig3}The total production cross-section of the process $ep\rightarrow \nu_4$X as a function of $m_{\nu_4}$ with the center of mass energy $\sqrt{s}$ = 3.74 TeV.}
\end{figure}
\section{Anomalous Single Production of the Fourth SM Generation Neutrino at \lowercase{ep} Colliders}

Single anomalous production of the fourth SM generation neutrino is considered at ep colliders with $\sqrt{s}$ = 1.4 TeV (QCD Explorer) and $\sqrt{s}$ = 3.74 TeV (Energy Frontier). The calculated cross-sections for $\nu_4$ are given for QCD Explorer and Energy Frontier ep collider in Fig. 2 and Fig. 3, respectively. We consider $ep\rightarrow\nu_4X\rightarrow \mu W X$ and $ep\rightarrow\nu_4X\rightarrow e W X$ processes as signatures of anomalous interactions of the fourth generation neutrino.  In order to extract the fourth generation neutrino signal and to suppress the background, we impose cuts on the eW invariant mass. Following {[}29{]}, cuts of $|m_{eW}-m_{\nu_4}|<25$ GeV for the mass range $m_{\nu_4} =100-1000$ GeV and $|m_{eW}-m_{\nu_4}|<50$ GeV for the mass range of $1-2.6$ TeV together with $p_{T}^{q, l}>10$ GeV are applied. In numerical calculations CTEQ6L  parton distribution functions are used {[}33{]}. The first process can be detected easily at ep colliders due to no background. Number of events for this process at $\sqrt{s} = 1.4$ TeV and $\sqrt{s} = 3.74$ TeV are presented in Table II and III, respectively. The computed signal and background cross-sections for the second process are given in Tables IV and V for $\sqrt{s} = 1.4$ TeV and $\sqrt{s} = 3.74$ TeV options, respectively. In the last two columns of these tables we present statistical significance (SS) values of the signal (evaluated from $SS=(\sigma_{S}/\sqrt{\sigma_B})\sqrt{L_{int}}$, where $L_{int}$ is the integrated luminosity of the collider).

\begin{table}
\caption{The cross section of signal and background of $ep\rightarrow\nu_4X\rightarrow e W X$ for $\sqrt{s}$ = 1.4 TeV, 
$(\kappa/\Lambda)$ = 1TeV$^{-1}$}
\begin{ruledtabular}
\begin{tabular}{ccccc}
& & &\multicolumn{2}{c} {$SS$}  \\
\cline{4-5}
$m_{\nu_4}$(GeV) & $\sigma_{S}$(pb) &$\sigma_{B}$(pb) & $L_{int}=1$ fb$^{-1}$ & $L_{int}=10$ fb$^{-1}$ \\ 
\colrule
200\hphantom{00} & \hphantom{0}0.201 & \hphantom{0} 0.560 & \hphantom{0}8.49 & 26.86 \\
300\hphantom{00} & \hphantom{0}0.148 & \hphantom{0} 0.293 & \hphantom{0}8.64 & 27.34 \\
400\hphantom{00} & \hphantom{0}0.106 & \hphantom{0}0.172  & \hphantom{0}8.08 & 25.56 \\
500\hphantom{00} & \hphantom{0}0.074 & \hphantom{0} 0.086 & \hphantom{0}7.98 & 25.23 \\
600\hphantom{00} & \hphantom{0}0.047 & \hphantom{0}0.049  & \hphantom{0}6.71 & 21.23 \\
700\hphantom{00} & \hphantom{0}0.027 & \hphantom{0} 0.025 & \hphantom{0}5.40 & 17.07 \\
800\hphantom{00} & \hphantom{0}0.014 & \hphantom{0}0.012  & \hphantom{0}4.04 & 12.78  \\
900\hphantom{00} & \hphantom{0}0.006 & \hphantom{0} 0.005 & \hphantom{0}2.64 & 8.34  \\
1000\hphantom{00} & \hphantom{0}0.002 & \hphantom{0}0.004  & \hphantom{0}1.10 & 3.48 \\ 
\end{tabular}
\end{ruledtabular}
\end{table}
\begin{table}
\caption{The cross section of signal and background of $ep\rightarrow\nu_4X\rightarrow e W X$ for $\sqrt{s}$ = 3.74 TeV, 
$(\kappa/\Lambda)$ = 1TeV$^{-1}$}
\begin{ruledtabular}
\begin{tabular}{ccccc}
& & &\multicolumn{2}{c} {$SS$}  \\
\cline{4-5}
$m_{\nu_4}$(GeV) & $\sigma_{S}$(pb) &$\sigma_{B}$(pb) & $L_{int}=100$ pb$^{-1}$ & $L_{int}=1$ fb$^{-1}$ \\ 
\colrule

200\hphantom{00} & \hphantom{0}0.743 & \hphantom{0} 1.605 & \hphantom{0}5.86 & 18.54 \\
300\hphantom{00} & \hphantom{0}0.603 & \hphantom{0} 0.954 & \hphantom{0}6.17 & 19.52 \\
400\hphantom{00} & \hphantom{0}0.504 & \hphantom{0}0.560  & \hphantom{0}6.73 & 21.30 \\
500\hphantom{00} & \hphantom{0}0.427 & \hphantom{0} 0.380 & \hphantom{0}6.93 & 21.90 \\
600\hphantom{00} & \hphantom{0}0.364 & \hphantom{0}0.280  & \hphantom{0}6.88 & 21.75 \\
700\hphantom{00} & \hphantom{0}0.307 & \hphantom{0} 0.209 & \hphantom{0}6.71 & 21.23 \\
800\hphantom{00} & \hphantom{0}0.259 & \hphantom{0}0.157  & \hphantom{0}6.53 & 20.67  \\
900\hphantom{00} & \hphantom{0}0.214 & \hphantom{0} 0.125 & \hphantom{0}6.05 & 19.14  \\
1000\hphantom{00} & \hphantom{0}0.201 & \hphantom{0}0.188  & \hphantom{0}4.63 & 14.66 \\ 
1200\hphantom{00} & \hphantom{0} 0.138 & \hphantom{0} 0.116 & \hphantom{0}4.05 & 12.81 \\
1400\hphantom{00} & \hphantom{0}0.088 & \hphantom{0}0.075  & \hphantom{0}3.21 & 10.16 \\
1600\hphantom{00} & \hphantom{0}0.053 & \hphantom{0}0.044  & \hphantom{0}2.52 & 8.00 \\
1800\hphantom{00} & \hphantom{0}0.029 & \hphantom{0}0.026  & \hphantom{0}1.80 & 5.69  \\
2000\hphantom{00} & \hphantom{0}0.015 & \hphantom{0}0.014  & \hphantom{0}1.27 & 4.01 \\ 
2200\hphantom{00} & \hphantom{0} 0.007 & \hphantom{0} 0.008 & \hphantom{0}0.80 & 2.47 \\
2400\hphantom{00} & \hphantom{0}0.003 & \hphantom{0}0.004  & \hphantom{0}0.50 & 1.50 \\
2600\hphantom{00} & \hphantom{0}0.001 & \hphantom{0}0.002  & \hphantom{0}0.20 & 0.71 \\
\end{tabular}
\end{ruledtabular}
\end{table}
\begin{figure}
\includegraphics{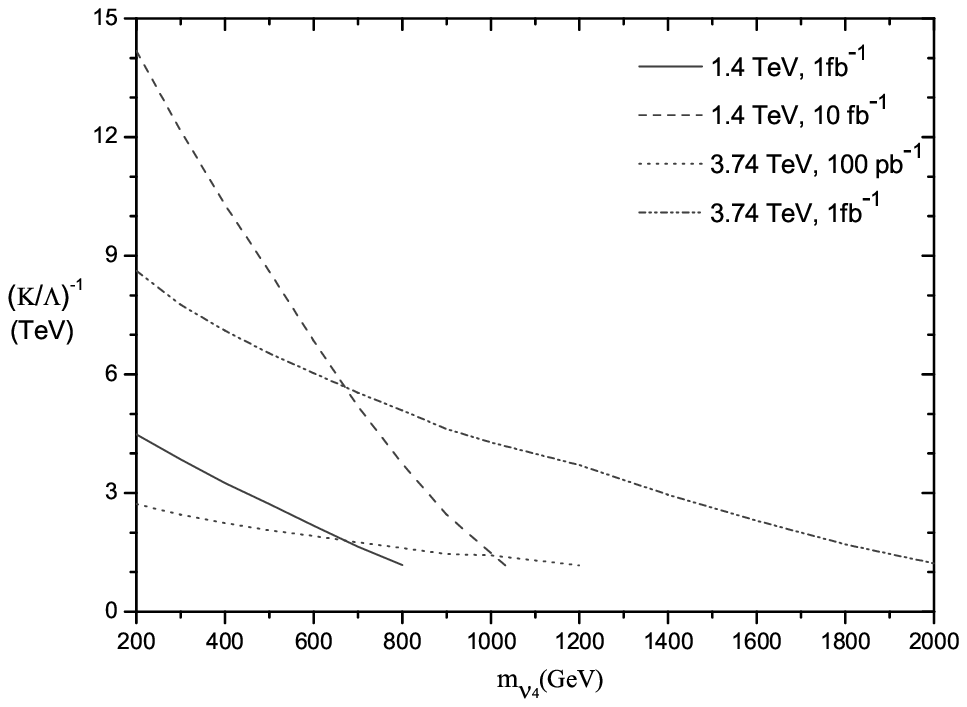}
\caption{\label{fig4}Achievable values of anomalous coupling strength as a function of the fourth generation neutrino mass for $ep\rightarrow\nu_4X\rightarrow \mu W X$ process.}
\end{figure}
\begin{figure}
\includegraphics{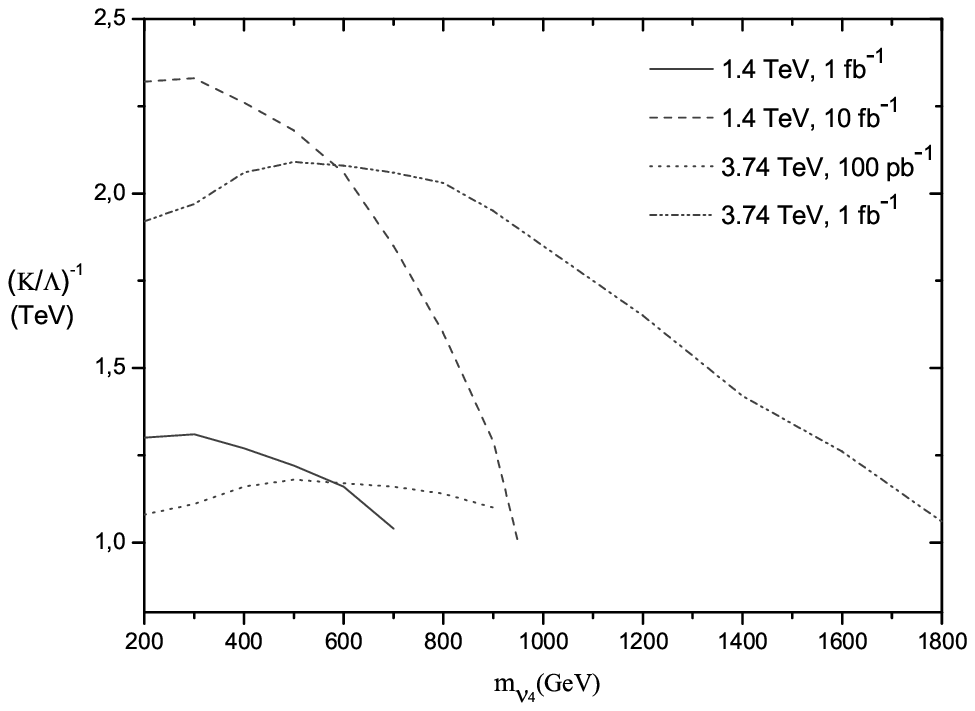}
\caption{\label{fig5}Achievable values of anomalous coupling strength as a function of the fourth generation neutrino mass for $ep\rightarrow\nu_4X\rightarrow e W X$ process.}
\end{figure}

Since the process $ep\rightarrow\nu_4X\rightarrow \mu W X$ has no SM background one can use 10 events as a discovery limit. As seen from Table II, QCD Explorer will reach $m_{\nu_4}= 850$ GeV (1100 GeV) with integrated luminosity of 1 fb$^{-1}$ (10 fb$^{-1}$) for $(\kappa/\Lambda) = 1$ TeV$^{-1}$. Corresponding limit for Energy Frontier is 1300 GeV (2100 GeV) with $L_{int}=100$ pb$^{-1}$ (1 fb$^{-1}$). Achievable values of anomalous coupling strength as a function of the fourth generation neutrino mass for process under consideration are shown in Fig. 4 for different ep collider options. One can see that   values as low as 0.077 TeV$^{-1}$ are reachable for $(\kappa/\Lambda)$.      

For the process $ep\rightarrow\nu_4X\rightarrow e W X$ we require $SS>5$ as a discovery criterion. It is seen from Tables IV and V, that QCD Explorer will cover masses of the fourth generation neutrino up to 750 GeV (950 GeV) with $L_{int}=1$ fb$^{-1}$ (10 fb$^{-1}$), whereas Energy Frontier ep collider will extend the mass region up to $m_{\nu_4}= 950$ GeV (1900 GeV) with $L_{int}=100$ pb$^{-1}$ (1 fb$^{-1}$). Figure 5 show that this channel is less promising than above one concerning achievable values of anomalous coupling strength. 

\section{Conclusion}
Combining results of studies on anomalous single production of the fourth SM generation charged lepton {[}28{]} and neutrino (this study) at future ep colliders we conclude that they have promising potential on the subject. For example, if $m_{\nu_4}= m_{l_4}= 500$ GeV and $(\kappa/\Lambda)$ = 1TeV$^{-1}$, the numbers of produced events are 740 for $ep\rightarrow\nu_4X\rightarrow e W X$ and 1100 for $ep\rightarrow l_4X\rightarrow e Z X$ at $\sqrt{s} = 1.4$ TeV with $L_{int}=10$ fb$^{-1}$. Finally, QCD Explorer cover $m_{\nu_4, l_4}< 1$ TeV, whereas Energy Frontier enlarge the region up to 2 TeV.   

\begin{acknowledgments}
This work is supported by TAEK and DPT with grant number DPT2006K-120470. 
\end{acknowledgments}

\end{document}